\documentclass[pre,preprint]{revtex4-1}
\usepackage{amsmath}
\usepackage{times}
\usepackage{graphicx}
\usepackage{color}
\usepackage{multirow}
\usepackage{hyperref}
\usepackage{url}
\usepackage{amssymb}
\usepackage{amsfonts}
\usepackage{indentfirst}
\usepackage{amsfonts}

\newcommand{\<}{\left<}
\renewcommand{\>}{\right>}
\newcommand{\ket}[1]{\left| #1 \>}
\newcommand{\bra}[1]{\< #1 \right|}
\newcommand{\braket}[2]{\< #1 | #2 \>}
\newcommand{\re}{\textup{Re}}

\newcommand{\be}{\begin{eqnarray}}
\newcommand{\ee}{\end{eqnarray}}

\begin{document}


\title{From synaptic interactions to collective dynamics in random neuronal networks models: Critical role of eigenvectors and transient behavior}

\author{E. Gudowska-Nowak$^\ddagger$}
\author{M.  A. Nowak$^\ddagger$\thanks{maciej.a.nowak@uj.edu.pl} }
\author{D.R. Chialvo$^{\dagger \#}$}
\author{J. K.  Ochab$^\ddagger$}
\author{W.  Tarnowski$^\ddagger$\thanks{wojciech.tarnowski@student.uj.edu.pl}}
\affiliation{$^\ddagger$Marian  Smoluchowski Institute of Physics and Mark Kac Complex Systems Research Center, Jagiellonian University, S. \L ojasiewicza 11,
PL 30-348 Krak\'ow, Poland.
}

\affiliation{$^{\dagger}$Center for Complex Systems \& Brain Sciences (CEMSC$^3$), Escuela de Ciencia y  Tecnolog\'ia,
 Universidad Nacional de San Mart\'{i}n, 1650 Buenos Aires, Argentina}
 \affiliation{$^{\#}$Consejo Nacional de Investigaciones Cient\'{i}ficas y Tecnol\'{o}gicas (CONICET),1650, Buenos Aires, Argentina.}

\begin{abstract} 
The study of neuronal interactions is currently at the center of several big collaborative neuroscience  projects (including the Human Connectome Project, the  Blue Brain Project, the  Brainome, etc.) which attempt to obtain a detailed map of the entire brain. Under certain constraints, mathematical theory can advance predictions of the expected neural dynamics based solely on the statistical properties of the synaptic interaction matrix. This work explores the application of free random variables to the study of large synaptic interaction matrices.
Besides recovering in a straightforward way known results on eigenspectra in types of models of neural networks proposed by Rajan and Abbott, we extend them to heavy-tailed distributions of interactions. 
More importantly, we derive analytically the behavior of eigenvector overlaps, which determine the stability of the spectra.
We observe that upon imposing the neuronal excitation/inhibition balance, despite the eigenvalues remaining unchanged, their stability dramatically decreases due to the strong non-orthogonality of associated eigenvectors.
It leads us to the conclusion that the understanding of the temporal evolution of asymmetric neural networks requires considering the entangled dynamics of both eigenvectors and eigenvalues, which might bear consequences for learning and memory processes in these models. 
Considering the success of free random variables theory in a wide variety of disciplines, we hope that the results presented here foster the additional application of these ideas in the area of brain sciences. 
\end{abstract}

\maketitle

\section*{Introduction}
Contemporary neuroscience focuses on detailed studies of the neuronal connections across the entire human brain.  Large scale collaborative efforts~\citep{BD1,BD2} including the BRAIN Initiative in the USA, Brainome in China, and the BlueBrain in the European Union were launched with the objective of mapping the connectivity of the entire brain at different resolutions. At a certain point a theory will be desperately needed to analyze these very large maps, describing the adjacency matrix of the brain. The work presented here is an attempt to enter into this uncharted and challenging territory. 

Under certain constraints, mathematical theory can advance predictions of the expected neural dynamics based solely on the statistical properties of their synaptic interaction matrix. In that sense randomly connected networks of neurons are one of the classical tools  of theoretical neuroscience. Only  recently  it was observed that the non-normality of the synaptic connectivity matrix (i.e., the matrix does not commute with its transpose)  has dramatic consequences for the temporal dynamics of stochastic equations,  which  can mimic  the dynamics of the network~\citep{HENNEQUIN,GANGULI,FUMAROLA}. In  particular, the work of Mart\'{i} et al.~\cite{MARTI} shows that increasing the symmetry of the connectivity leads to a systematic slowing-down of the dynamics and {\it vice versa},  decreasing the symmetry of the matrix leads to the  speeding  of the dynamics. This asymmetry  not only forces matrices to have complex spectra (which challenges  several traditional tools of random matrix theory), but more importantly, its study sheds new light on the role of the Bell-Steinberger~\citep{BS} matrix of overlaps  between the left and right eigenvectors of the connectivity matrix. 

Contemporarily,  the pivotal role of overlaps is understood in the simplest case of the spectral dynamics of the complex Ginibre matrix - either in Smoluchowski-Fokker-Planck formalism~\citep{OURPRL, OURNPB} or in Langevin formalism~\citep{DUBACH,GRWAR}, following the pioneering paper~\citep{CHALMEH,MEHLIGCHALKER}. The effects of the overlaps of the Ginibre matrix  for the temporal autocorrelation function of randomly connected networks was recently addressed analytically~\citep{MARTI},  confirming the numerical simulations in the weakly coupled regime of synaptic models. 

In this paper we study the non-normality aspects of the popular model with excitatory-inhibitory structure~\citep{EXPBALANCED,EXPBALANCED2,EXPBALANCED3}, proposed by~\citet{ABBOTTRAJAN}. An important ingredient of this model is the  introduction of the balance condition, which  stabilizes the fluctuating spectra of the network. Later, the numerical study of the full non-linear dynamics in the Rajan-Abbott model~\citep{WAINRIB} has shown the emergence of a transition leading to synchronized (stationary or periodic) states. This phenomenon cannot be explained solely by the  spectral features of the connectivity matrix, which motivates our study of missing non-spectral properties of non-normal networks, such as sensitivity to perturbations and transient dynamics induced by the non-orthogonality of eigenvectors. Recently, it was also hypothesized that the non-normality is universal in real complex networks~\citep{ASLLANI}.

Free random variables (hereafter FRV) theory is a relatively young mathematical theory, originating from the works of Voiculescu~\citep{Voiculescu}.  Partly due to the connection with large random matrices, it made in last decade a huge impact on physics~\citep{PHYSICS}, statistical inference~\citep{INFERENCE,INFERENCE2}, engineering of ICT technologies~\citep{MIMO} and finances~\citep{FINANCES,FINANCES2,FINANCES3,FINANCES4}.  In brief, FRV can be viewed as a non-commutative probability theory  for Big Data problems, where the information is hidden in statistical properties of eigenvalues and eigenvectors. As such, it is  ideally suited for disentangling signals from noise in various kinds of complex systems. Another advantage comes form the fact, that at the operational level the formalism is simple and powerful, allowing very often to  get results  on the basis of ``back-of-the-envelope" calculations.  

From this perspective, it is rather bewildering, that FRV so far has not been broadly applied to the most challenging  complex problem of understanding the brain. Thus, in this paper we consider FRV applications to understand the neuronal networks as represented by the synaptic strength matrix.  
A direct application of FRV  not only allows us to recover in a straightforward way well known  results from the literature~\citep{ABBOTTRAJAN}, but also to address quantitatively such issues as the stability of the network with respect to perturbation and extension the existing formalisms for the heavy-tailed distributions. 

The paper is organized as follows. 
In Sec. \ref{Sec:Nonnorm} we discuss two important effects caused by the non-orthogonality of eigenvectors of non-normal matrices, namely high sensitivity of the spectra and the transient behavior of the linearized dynamics. We briefly describe free probability theory in Section~\ref{Sec:FRV}, showing how it allows one to calculate the spectral density and gives an access to the eigenvector non-orthogonality.
In Section~\ref{sec:networks} we reframe the model introduced by  Rajan  and Abbott  in this language. 
Applying the theoretical toolbox explained in Appendices~A-C we recover and generalize their main results for the unbalanced network. In doing so, we uncover the analytic formulas for the one-point eigenvector correlation function for this model, which is crucial  for the determining its stability.
Since FRV  work also in the case of heavy-tailed distributions~\citep{HEAVY}, we present results for the spectra and eigenvectors of the Rajan-Abbott  model adapted for the case of  Cauchy noise.
We successfully  confirm our analytic predictions with numerical simulations.

Further, in Section~\ref{Sec:Nonnormality}, we show explicitly that the excitation/inhibition balance condition not only  tames the spectral outliers, but also exerts dramatic effects on the non-orthogonality of eigenvectors, increasing the networks' eigenvalue condition number by several orders of magnitude.
Section~\ref{sec:summup} closes the paper with a summary of the main results and their implications.  It also outlines main promising directions for further studies using the presented formalism.

\section{Non-normality of synaptic interactions in neural networks} \label{Sec:Nonnorm}

Adjacency matrices of directed networks and synaptic strength matrices are non-normal. This influences not only their spectra, as the eigenvalues can be complex, but also has a strong effect on the eigenvectors. A diagonalizable non-normal matrix possesses two eigenvectors: left and right for each eigenvalue. They satisfy the eigenproblems
\begin{equation}
\bra{L_i}X=\bra{L_i}\lambda_i,\qquad  X\ket{R_i}=\lambda_i\ket{R_i}.
\end{equation}
We use here physicists' ``bra-ket notation'', where $\ket{R_i}$ is a column  and $\bra{L_i}$ is a row vector. The scalar product is denoted as $\braket{L_i}{R_j}$ and we define the conjugated left vector $\ket{L_i}={(\bra{L_i})}^{\dagger}$.

Eigenvectors are normalized to $\braket{L_i}{R_j}=\delta_{ij}$, but they are not orthogonal among themselves $\braket{R_i}{R_j}\neq\delta_{ij}\neq\braket{L_i}{L_j}$. Chalker and Mehlig  introduced a matrix of scalar products of eigenvectors~\citep{CHALMEH,MEHLIGCHALKER}
\begin{equation}
O_{ij}=\braket{L_i}{L_j}\braket{R_j}{R_i}.
\end{equation}
Below we describe two phenomena important in neural networks, in which the non-orthogonality of eigenvectors captured in the matrix of overlaps plays an essential role.

\subsection{Perturbations of a network}

Considering the perturbation of the matrix $X$ by some $\epsilon P$, the change of the spectrum in the first order in $\epsilon$ reads
\be
\delta \lambda_i=\epsilon \bra{L_i}P\ket{R_i} \le  \epsilon \sqrt{\braket{L_i}{L_i} \braket{R_i}{R_i}} ||P||_F.
\label{triangle}
\ee
The inequality follows from the Cauchy inequality and  $||P||_F$ denotes the Frobenius norm  $||P||_F^2={\rm Tr} PP^{\dagger}$.
This inequality is saturated (equality holds) by the rank one  Wilkinson matrix $P=\ket{L_i}\bra{R_i}$. The inequality above shows that spectra of networks represented by non-normal matrices are more sensitive to changes in their connectivity. This enhanced sensitivity is driven by the non-orthogonality of eigenvectors. The quantity $\kappa(\lambda_i)=\sqrt{O_{ii}}$ is known in the numerical analysis community as the eigenvalue condition number~\citep{Wilkinson,Trefethen}.

\subsection{Eigenvector non-orthogonality in transient dynamics}
Stability analysis and the linear response of the dynamic systems with respect to external perturbations are among the most popular methods of describing complex systems~\citep{GUCKENHEIMER}. 
Let us consider dynamics obtained from the linearization of the system in the vicinity of a fixed point 
\be
\frac{d}{dt} \ket{\psi}=(-\mu +X)\ket{\psi} +\ket{\xi(t)}. \label{eq:LinearDynamics}
\ee 
Here $\xi$ represents the external drive and $\mu$ ensures stability in the absence of coupling ($X$) between components. In the context of neural networks, $\mu$ represents the current leakage due to membrane capacitance~\citep{SOMPOLINSKY}. 
Choosing it as a ``spike'' $\ket{\xi(t)}=\delta(t)\ket{\psi(0)}$ or, equivalently, choosing an initial condition $\ket{\psi(0)}$,  we formally  solve the system for $t>0$
\be
\ket{\psi(t)} =\exp [(X-\mu)t] \ket{\psi(0)}. \label{eq:FormalSol}
\ee

The long-time dynamics is governed by the eigenvalue with the largest real part. However, if $X$ is non-normal, this analysis is incomplete. The behavior of the linearized dynamics can be drastically different at its early stage. In particular, the system may initially move away from the fixed point. This sometimes invalidates the linear approximation and renders the fixed point unstable, even though the linearized dynamics predicts stability.  

To describe this transient dynamics, we consider the squared Euclidean distance from the fixed point, which is the squared norm of the solution (\ref{eq:FormalSol})
\be
D(t)=\braket{\psi(t)}{\psi(t)}&=& e^{-2\mu t} \bra{\psi(0)}e^{X^{\dagger}t} e^{Xt}\ket{\psi(0)} \nonumber \\
&=&\sum_{i,j=1}^N\braket{\psi(0)}{L_i}\braket{R_i}{R_j}\braket{L_j}{\psi(0)}e^{-2\mu t +t(\bar{\lambda}_i+\lambda_j)}. \label{eq:SquaredNorm}
\ee
If we consider  $\ket{\psi(0)}$ as a particular vector of unit norm, averaging over all directions uniformly distributed on the hypersphere (real or complex) $||\psi(0)||^2=1$  leads to 
\be
\bar{D}(t)=e^{-2\mu t} \frac{1}{N} {\rm Tr} e^{X^{\dagger}t }e^{Xt}=e^{-2\mu t} \frac{1}{N}\sum_{ij}e^{ t(\lambda_i+\bar{\lambda}_j)}O_{ij} .
\label{unitary}
\ee

We see that all elements of the overlaps of left and right eigenvectors drive the behavior of the squared distance. First, they enhance the contributions of the eigenmodes, which is responsible for amplification of the response to the external driving. Second, since the matrix is not diagonal, they couple different eigenmodes during the evolution. This results in an interference between eigenmodes, which is reflected as an oscillatory behavior of the squared norm of the solution (see also Fig~\ref{Fig:SquaredNorm}). Note that for normal matrices, such effects do not exist, since left and right vectors are orthogonal and the ``coupling matrix" is an identity. 
Recently, the transient growth was proposed as an amplification mechanism of neural signals~\citep{MURPHY,HENNEQUIN,HENNEQUIN2}. We also remark here that even in the systems in which the average trajectory is not amplified, one can still observe transient trajectories, provided that the initial condition is chosen from the subspace spanned by the eigenvectors of $e^{X^{\dagger}t}e^{Xt}$ to eigenvalues greater than 1~\citep{BONDANELLI}.

Usually the matrix $X$ is modeled as random. We remark that the averaging over all initial conditions is equivalent to fixing an initial vector $\ket{\psi(0)}$ and averaging over the vectors $U\ket{\psi(0)}$, where $U$ is uniformly distributed (according to the Haar measure) on the orthogonal (unitary) group. This implies that the average over initial conditions is already included when averaging over randomness in $X$ when its probability density function is invariant under orthogonal  (unitary) transformations, $P(X)=P(UXU^{\dagger})$. Even though the matrix $X$ may not admit this invariance, the averaging over initial conditions is equivalent to rotating the matrix $e^{X^{\dagger}t}e^{Xt}\to U^{\dagger}e^{X^{\dagger}t}e^{Xt}U$, thus acting as if $X$ were invariant. Although biologically plausible models break the unitary invariance of the synaptic connectivity matrix, the above argument and Eq~\eqref{unitary} apply to a broad class of models.

\section{Theory of Free Random Variables}
\label{Sec:FRV}
\subsection{Spectral density and eigenvector correlations}

Unitarily (and orthogonally) invariant random matrices in the large size limit are described by free probability. Its power relies on the easiness of obtaining analytical formulas, which are very good approximations even for  relatively small  matrix size. 

An important class of matrices, the so-called bi-unitarily invariant, which generalizes the Gaussian distribution (described in Sec. \ref{Sec:ExampleGG}) is important in models of neural networks. In this class, the unitary symmetry of the distribution is enhanced to $P(UXV)=P(X)$ for $U,V$ independent unitary matrices, hence the name.
Despite the fact that they are genuinely non-Hermitian, due to enhanced symmetry the spectral problem is effectively one-dimensional, because the spectrum is rotationally invariant on the complex plane.
 In this case, a powerful result  holds in FRV, known as the Haagerup-Larsen theorem~\citep{HL}. It states that the radial cumulative distribution function $F(r)=\int_0^{r} 2\pi \rho(r')r'dr'$, of the ensemble $X$  can be inferred from the simple functional equation
\be
S_{X^{\dagger}X}(F(r)-1)=\frac{1}{r^2}
\label{HLspectrum}
\ee
where $S_X(z)$ is the so-called S-transform for the ensemble $X$.
In Appendix A we explain the probabilistic interpretation of $S$ and we provide a simple example. 
Spectra of bi-unitarily invariant ensembles in large $N$ limit are supported on either a disc or an annulus, a phenomenon dubbed ``the single ring theorem"~\citep{FEINZEE,FEINZEESCAR}. The inner radius of the spectrum is deduced from the condition $F(r_{in} )=0$, while the outer one is given by $F(r_{out})=1$.

The applicability of free probability to non-Hermitian matrices is not limited to spectra only. It gives also access to the averages of the overlap matrix conditioned on eigenvalues. The one-point function
\begin{equation}
O(z)=\frac{1}{N^2} \left< \sum_{i=1}^{N} \delta^{(2)}(z-\lambda_i) \braket{L_i}{L_i} \braket{R_i}{R_i}\right>,
\end{equation}
associated with the diagonal elements of the overlap matrix can be calculated for any type of unitarily invariant probability~\citep{JNEGVEC}. For bi-unitarily invariant ensembles it takes the remarkably simple form~\citep{BNST}
\be 
O(r)= \frac{1}{\pi r^2} F(r)(1-F(r)).
\label{HLeigenvector}
\ee

 The ratio of the one-point correlation function and the spectral density gives the conditional expectation of the squared eigenvalue condition number~\citep{BNST}
\begin{equation}
\mathbb{E}\left(\kappa^{2}(\lambda_i)|r=|\lambda_i|\right)=\frac{N O(r)}{\rho(r)}.
\end{equation} 

Recently, the two-point function associated with off-diagonal elements of the overlap matrix has become accessible within free probability~\citep{PROBING}.

\subsection{Example: Ginibre-Girko ensemble}
\label{Sec:ExampleGG}
We conclude this section with an example of the above construction by considering the so called Ginibre-Girko matrix $G$, the entries of which are independently taken  from the real/complex Gaussian distribution with zero mean and $1/N$ variance. Such a case was already considered in the model of randomly connected neural networks by \citet{SOMPOLINSKY}.

According to Eq~(\ref{HLspectrum}), we need the S-transform for $G^{\dagger}G$. This matrix belongs to the Wishart ensemble~\citep{WISHART,WISHART2}. Its $S$-transform reads $S_{G^{\dagger} G}(z)=\frac{1}{1+z}$ (see Appendix~A). 
 This completes the calculation, since now replacing $z \rightarrow F(r)-1$ and using  Eq~(\ref{HLspectrum}) we get
\be
F(r)=r^2.
\ee
 The spectrum is therefore uniform,  $\rho(r)=\frac{1}{2\pi r} \frac{dF(r)}{dr}=\frac{1}{\pi}$, on the unit disc ($F(r_{in})=0$, $F(r_{out})=1$), reproducing the Ginibre-Girko result. The eigenvector  correlator comes from Eq~(\ref{HLeigenvector}), $O(r)=\frac{1}{\pi}(1-r^2)$, in agreement with~\citet{CHALMEH}, where it was calculated using much more laborious techniques. 
In the next section, we show that the same computational simplicity is preserved when considering the ensembles taking into account physiological restrictions imposed on  the neural networks models.

\section{Reframing Rajan-Abbott model}
\label{sec:networks}

The strength of synapses between all pairs of $N$ neurons in a network is represented by the weighted adjacency (synaptic) matrix. Contrary to the Ginibre matrices, the structure of its elements is more complicated. In the minimal model~\citep{ABBOTTRAJAN}, there are two kinds of neurons with a fraction $f_E N$ representing excitatory ($E$), and $f_I N=(1-f_E) N$ the remaining inhibitory ($I$) neurons. Their strengths are sampled from Gaussian ensembles, with means $\mu_i$ and variances $\sigma_i^2/N$, where $i=I,E$. The matricial representation of the synaptic strength matrix reads therefore $X=M+W$. Here the deterministic matrix $M$ represents the average synaptic activity. In this model it is a rank one matrix with identical rows, each containing $f_E N$ consecutive means $\mu_E$ and followed by $f_I N$ consecutive means $\mu_I$. The random part $W$ models variability across the population. It is assumed to be of the form $W=G\Lambda$, where $G$ is the Girko-Ginibre matrix and $\Lambda$ is diagonal with its first $f_EN$ elements equal to $\sigma_E$ and last $f_IN$ ones equal to $\sigma_I$.

\begin{figure}[!]
 \centering
\includegraphics[width=0.85\textwidth]{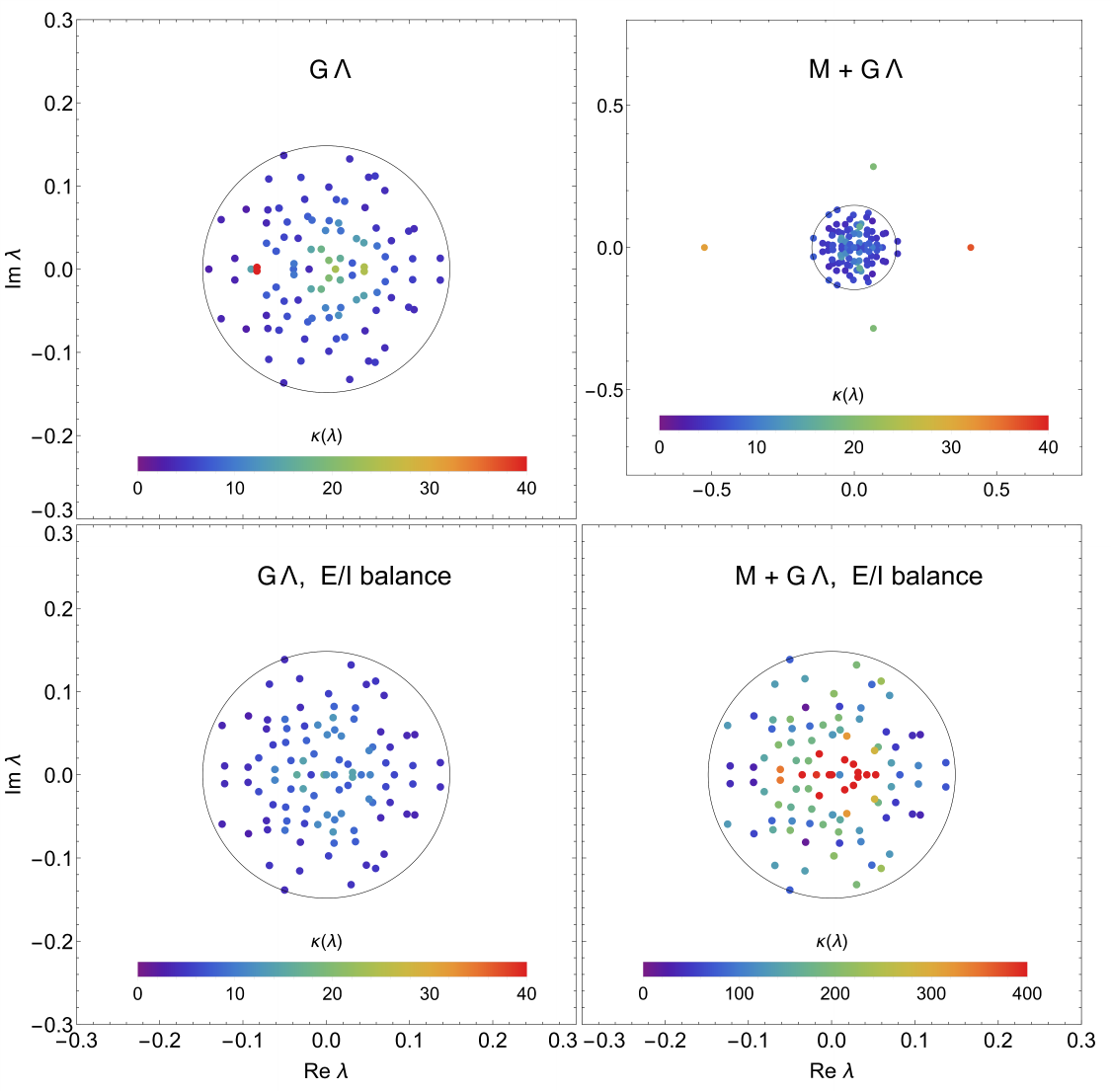} 
 \caption{Eigenvalues and their condition numbers of the matrix of variances $G\Lambda$ (top left), the Rajan-Abbott neural network model (top right), matrix of variances with the E/I balance imposed (bottom left) and the Rajan-Abbott model with E/I balance (bottom right).
We observed in many realizations that the outliers of the unconstrained Rajan-Abbott model have a higher condition number than the average of eigenvalues within the circle.
The spectra in the panels on the left differ only slightly.  The eigenvalues presented in bottom panels are exactly the same, but the presence of a highly non-normal matrix $M$ causes the eigenvalues on the bottom right to be conditioned much more poorly: note the tenfold ($\sqrt{N}$, as predicted by Eq~(\ref{eq:CorrModified})) broader scale in that panel.
The same realization of the Gaussian matrix $G$ was taken for all plots. We used parameters $\sigma_I=0.3$, $\sigma_E=0.1$, $f_I=0.15$, $f_E=0.85$, $\mu_I=0.85$, $\mu_E=0.15$, and matrix size $N=100$.\label{Fig:SpectraComparison} 
}
\end{figure}

Several studies~\citep{BALANCE,BALANCE2,EXPBALANCED3} show that the amount of excitation and inhibition of a neuron is the same (the so-called E/I balance) even on the scale of few milliseconds~\citep{EXPBALANCED,EXPBALANCED2}. 
To incorporate this fact in the model, the balance condition is imposed on two levels. The global condition $f_E\mu_E + f_I\mu_I=0$  means that neurons are balanced on average. This forces the last non-zero eigenvalue of $M$ to vanish.
Even in the case of a null spectrum of $M$, its non-normal character causes the eigenvalues of  $M+G\Lambda$ differ much from that of $G\Lambda$. As a result a few eigenvalues lie far beyond the spectrum of $G\Lambda$~\citep{ABBOTTRAJAN,TAO}, see Fig~\ref{Fig:SpectraComparison}. 
 
The local E/I balance is imposed on this model by demanding that the sum of strengths coupled independently to each neuron vanishes.
Mathematically, we subtract the $1/N$ of a sum of each row from any element in that row. As a consequence, the elements within each row sum to zero.
This condition brings the outliers back to the disc of radius $R=\sqrt{f_I\sigma_I^2+f_E\sigma_E^2}$ -- now the spectra of $W$ and $M+W$ are identical~\citep{ABBOTTRAJAN}, see also Fig~\ref{Fig:SpectraComparison}.
Whenever we indicate E/I balance in the figures, we assume such local balance.

\subsection{Rajan-Abbott results from FRV}
\label{sec:ARresult}

Having known that the E/I balance causes the spectrum to be insensitive to the matrix of average strengths $M$, we consider a more general model of $m$ types of neurons, each with multiplicity $f_kN$ and the synaptic strength variance $\sigma_k^2/N$. The random part of the synaptic strength matrix can be written as $W=G\Lambda$, where $G$ is a Ginibre-Girko matrix as before, while $\Lambda$ is  diagonal with a generic structure  $\textup{diag}(\sigma_1 \mathbf{1}_{f_1N} , ..., \sigma_m \mathbf{1}_{f_mN})$. The multiplicities are normalized as 
$\sum_{i=1}^m f_i=1$. In Appendix~B, using free probability, we obtain the algebraic equation for the radial cumulative distribution function $F(r)$ 
\be
 1=\sum_{i=1}^{m} \frac{f_i\sigma_i^2}{r^2-\sigma_i^2(F(r)-1)}.
 \label{finalXXXX}
 \ee
Explicit solutions exist for $m=2,3,4$ types of neurons, corresponding to the quadratic, cubic, or quartic algebraic equation for $F(r)$, but other cases are easily tractable numerically. The case solved by Rajan and Abbott  corresponds to the quadratic equation.  Solution (\ref{finalXXXX}) is also equivalent to the diagrammatic construction of \citet{WEI}, but more explicit. 
 The spectrum is always confined within the disc of radius $r_{out}^2=\sum_{i=1}^{m}  f_i \sigma_i^2$, as visible from the condition $F(r_{out})=1$.
 
We will argue in Sec.~\ref{Sec:Nonnormality} that the presence of the deterministic matrix $M$ and the balance condition exert a dramatic effect on the eigenvectors of the synaptic strength matrix. Knowing $F(r)$, free probability allows us to calculate via Eq~(\ref{HLeigenvector}) also the eigenvector correlation function $O(r)$ for its random part $W$.
In the case of the minimal model considered by Rajan and Abbott, it reads explicitly
\be
O_W(r)=\frac{1}{2\pi \sigma_E^4\sigma_I^4}\left( (f_I-f_E)\sigma_I^2\sigma_E^2(\sigma_E^2\sigma_I^2)-r^2(\sigma_E^4+\sigma_I^4)+(\sigma_E^2+\sigma_I^2)\sqrt{K}\right), \label{eq:RAcorrelator}
\ee
where 
\begin{equation}
K=r^4 {(\sigma_E^2-\sigma_I^2)}^2+\sigma_I^4\sigma_E^4+2r^2(f_E-f_I)\sigma_E^2\sigma_I^2(\sigma_E^2-\sigma_I^2).
\end{equation}
This result is inaccessible within the framework of~\citet{WEI}.

\subsection{Heavy-tailed noise}
\label{sec:cauchy}

Cauchy noise,  belonging to the regime of Lévy stable distributions, is used here as the simplest mechanism to mimic the non-Gaussianity  of the realistic  synaptic matrices. Since learning rules could change the initial random network structure into a small world network~\citep{WATTS,Yu2008,Downes2012,Pastore2018} by dynamic modification of  synaptic weights, the possibility of obtaining analytic benchmarks for heavy-tailed distributions  is appealing. 
Spatial and temporal Lévy processes are omnipresent in biological time series, but the fact that they  do not possess  finite moments invalidates several standard tools of statistical analysis.  In the case of  matrices exhibiting heavy-tailed distributions of elements, the underlying mathematical structure is  quite involved~\citep{BOUCHAUDCIZEAU,BURJURNOWLEV}. Here, for simplicity, we focus  on the Cauchy matrix distribution, given by the probaiblity density function
$
P(X)\sim \det(XX^{\dagger}+1)^{-2N}.
$

Application of FRV techniques to the spectral Cauchy distribution leads (see Appendix~C)  to the simple result 
\be
 \rho(r)&=&\frac{1}{2\pi r}\frac{dF(r)}{dr}=\frac{1}{\pi} \sum_{i=1}^{m} \frac{f_i\sigma_i^2}{{(r^2+\sigma_i^2)}^2}, \\
 O(r) &=& \frac{1}{\pi r^2} F(r)(1-F(r))= \frac{1}{\pi} \sum_{i=1}^{m}\frac{f_i}{r^2+\sigma_i^2} \sum_{j=1}^{m} \frac{f_j\sigma_j^2}{r^2+\sigma_j^2}.
 \label{Cauchyresults}
 \ee
 In this case, the spectrum spreads over the whole complex plane, reflecting the large fluctuation of Lévy type noise.
 In the case of more realistic Lévy noise, one loses the simple analytic structure presented above, but the formalism stays -- the resulting equations are usually of transcendental type, but can be easily solved numerically.  
 
 We also remark that in models with heavy-tailed randomness the Dale's principle cannot be tightly satisfied in this model. Irrespective of the mean of this distribution, there is always a non-negligible probability of obtaining the value with the opposite sign as the mean because of infinite variance of such distributions.

\section{Non-normality in the Rajan-Abbott model} 
\label{Sec:Nonnormality}

Below we argue that imposing the E/I balance not only confines the eigenvalues to a disc, but -- more importantly -- induces a very strong non-orthogonality of eigenvectors. This in turn causes the spectra to be highly sensitive to perturbations and strengthen the transient effects. 

Let us assume that the matrix $W$ is diagonalizable. If we denote $\ket{u}={(1,1,\ldots,1)}^T$, the E/I balance is equivalent to the fact that $\ket{u}$ is the right eigenvector of $W$ to the eigenvalue $\lambda_1=0$. Let $\bra{L_1}$ be the left eigenvector to this eigenvalue. For brevity we also denote $\bra{m}=(\underbrace{\mu_1,\ldots,\mu_1}_{f_1N\,\, \textup{times}},\ldots,\underbrace{\mu_m,\ldots,\mu_m}_{f_mN\,\, \textup{times}})$, which allow us to write $M=\ket{u}\bra{m}$. The spectral decomposition of $W$ reads
\begin{equation}
W=0 \cdot \ket{u}\bra{L_1}+\sum_{j=2}^{N}\ket{R_j}\lambda_j \bra{L_j}.
\end{equation}
Since $\braket{m}{u}=0$, $\bra{m}$ has a decomposition into the left eigenvectors of $W$, except for $\bra{L_1}$, thus $\bra{m}=\sum_{j=2}^{N}\bra{L_j}\alpha_j$ with $\alpha_{j}=\braket{m}{R_j}$. Hence, the total synaptic strength matrix is decomposed as
\begin{equation}
M+W=0\cdot \ket{u}\bra{L_1}+\sum_{j=2}^{N}\left(\ket{R_j}+\frac{\alpha_j}{\lambda_j}\ket{u}\right) \lambda_j\bra{L_j}.
\end{equation}
We constructed explicitly the eigenvectors of the synaptic strength matrix. The left eigenvectors are not altered when $M$ is taken into consideration due to the E/I balance. The bi-orthogonality condition $\braket{L_i}{R_j}=\delta_{ij}$ leaves freedom of rescaling each pair of eigenvectors by a non-zero complex number $\ket{R_j}\to c_j\ket{R_j}$ and $\bra{L_j}\to\bra{L_j} c_j^{-1}$. These transformations allow us to set the length of left eigenvectors $\braket{L_j}{L_j}=1$. The diagonal elements of the overlap matrix in the presence of the matrix $M$ and E/I balance read
\begin{equation}
O'_{jj}=O_{jj}+2\re \left(\frac{\braket{m}{R_j}\braket{R_j}{u}}{\lambda_j}\right)+N\frac{|\braket{m}{R_j}|^2}{|\lambda_j|^2}, \label{eq:CorrModified}
\end{equation}
where we have used $\braket{u}{u}=N$ and denoted $O_{jj}$ the overlap matrix in the absence of deterministic weights $M$. Note that $O_{jj}$ also grows linearly with $N$. This shows that the condition numbers grow with the size of a matrix and the effect of the matrix of averages is stronger for eigenvalues close to the origin.

Analogous reasoning for the full overlap matrix leads to the conclusion that all its elements $O_{ij}$ for $i,j\geq 2$ are affected by the E/I balance and the deterministic matrix. The dominant term in large $N$ is given by
\begin{equation}
O'_{ij}-O_{ij}\sim N\braket{L_i}{L_j}\frac{\braket{m}{R_i}\braket{R_j}{m}}{\lambda_i\bar{\lambda_j}}. \label{eq:OffDiag}
\end{equation}

\begin{figure}[!] 
\begin{center}
\includegraphics[width=1.0\textwidth]{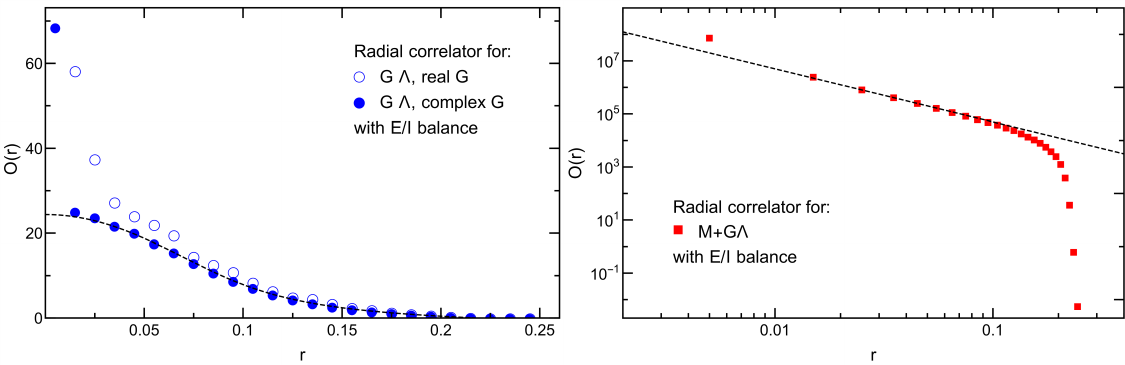} 
\end{center}
\caption{(left) The eigenvector correlation function for the matrix of variances $G\Lambda$ with the E/I balance imposed. The random matrix $G$ was generated from the complex and real Ginibre ensembles. The dashed line presents the analytical solutions from FRV. Numerical results (circles) were obtained by diagonalizing 1500 matrices of size $N=1000$. The discrepancies for real matrices come from the real eigenvalues. The fluctuations of the diagonal overlaps associated with them are so strong that the mean of their distribution does not exist~\citep{FYODOROV}.
(right, log-log scale) Eigenvector correlator of $M+G\Lambda$, where $G$ is complex Ginibre. The solid line presents the power-law, $O(r)\sim r^{-2}$, predicted by~Eq~(\ref{eq:CorrModified}) for small $r$. In both pictures we took parameters $\sigma_I=0.4$, $\sigma_E=0.1$, $f_I=0.25$, $f_E=0.75$. For the picture on the right we also set $\mu_E=0.25$, $\mu_I=0.75$. 
\label{Fig:corr_FRV}}
\end{figure}

To study the statistics of the eigenvalue condition numbers, we performed numerical simulations by diagonalizing matrices, the random part of which was generated from either real or complex Ginibre ensemble. The eigenvector correlation function is juxtaposed with (\ref{eq:RAcorrelator}) from free probability, see Fig~\ref{Fig:corr_FRV}. The presence of the matrix $M$ and the E/I balance is manifested in the scaling $O(r)\sim r^{-2}$ for small $r$, as observed in Fig~\ref{Fig:corr_FRV}, in accordance with Eq~(\ref{eq:CorrModified}).

There is a visible mismatch between numerics for real matrices and the results from free probability, particularly evident for eigenvalues with small moduli. This fact is explained in the light of the recent result of~\citet{FYODOROV}, who showed that the distribution of the overlap for Gaussian matrices is heavy-tailed. This distribution conditioned on real eigenvalues of the real Ginibre ensemble is so fat-tailed that even the mean does not exist, thus $O(z)$ can be considered only outside the real axis. Being aware of this fact, we have performed further simulations only for complex matrices which do not suffer from this problem. 

\begin{figure}[h]
\includegraphics[width=\textwidth]{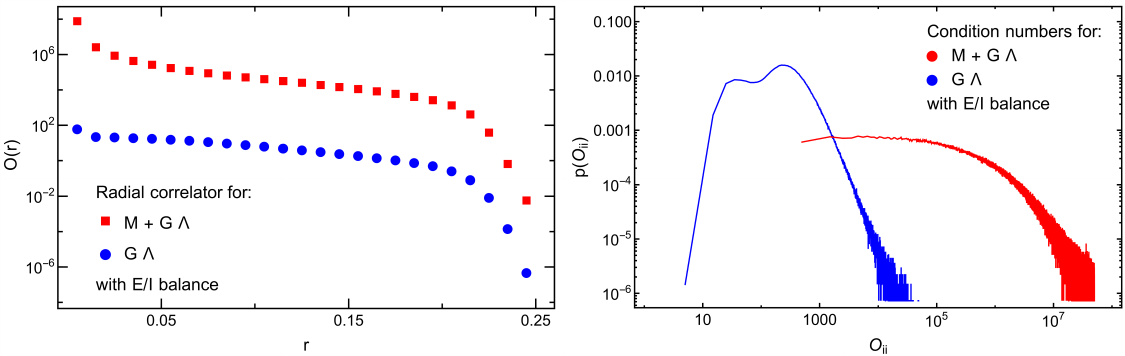}
\caption{(left, log scale) The eigenvector correlation function of the synaptic strength matrices with the E/I balance condition. Despite unchanged spectra, the squared condition numbers differ significantly.
(right, log-log scale) The distribution of the squared eigenvalue condition numbers of the eigenvalues of the synaptic strength matrix with and without the matrix $M$. 
\label{Fig:CorrJuxtaposition}
}
\end{figure}

We studied the effect of the deterministic matrix $M$ by juxtaposing the eigenvector correlation function in Fig~\ref{Fig:CorrJuxtaposition} and noticed the significant increase in its magnitude. This enhancement of non-normality is visible not only on the level of the mean value, but also on the full distribution of the overlap (see Fig~\ref{Fig:CorrJuxtaposition} (right)).

\begin{figure}[h]
\includegraphics[width=\textwidth]{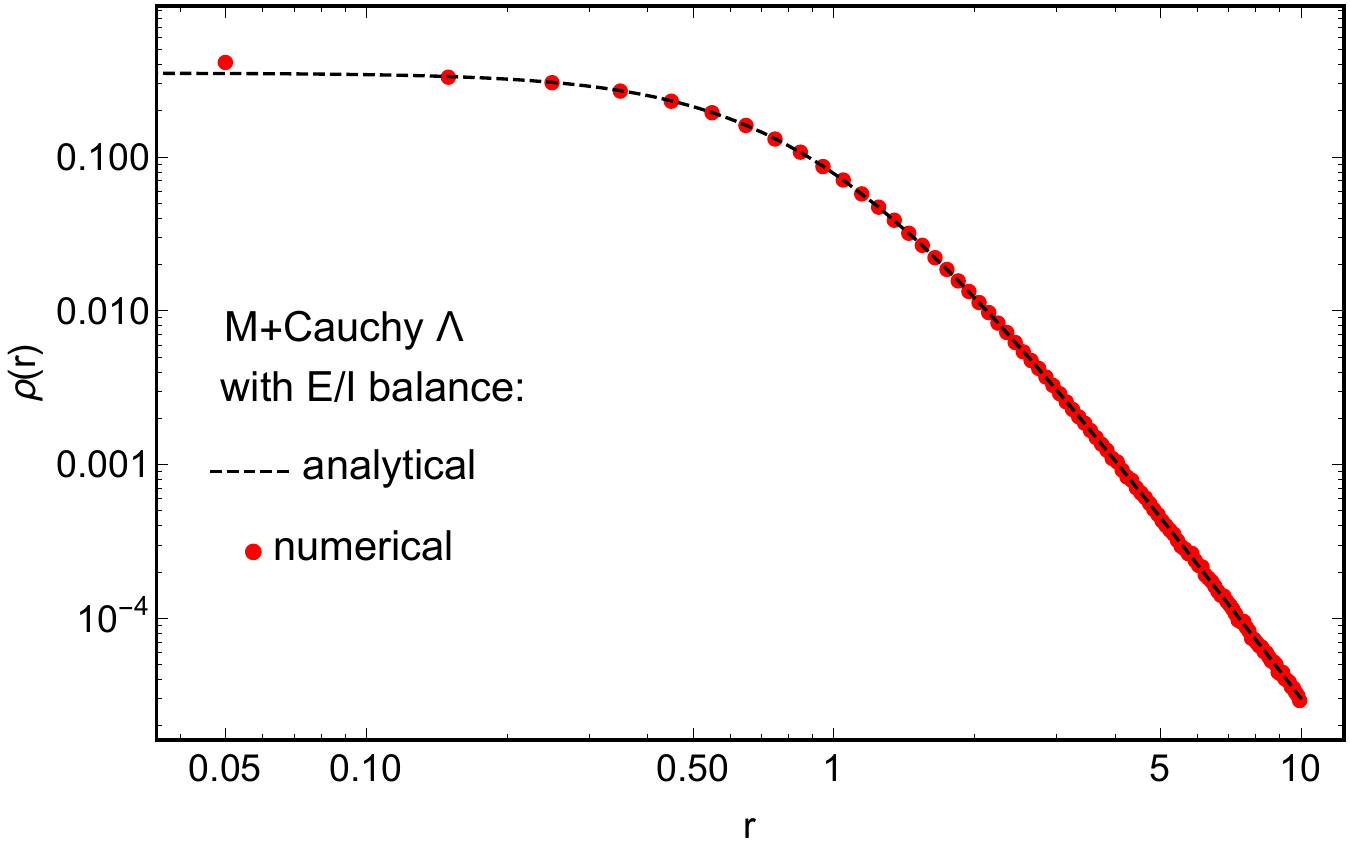}  
\caption{A cross-check of the numerical results with the analytical prediction of the spectral density for the Cauchy synaptic matrix on a log-log scale.
\label{Fig:FRVCauchy}
}
\end{figure}

\begin{figure}[h]
\includegraphics[width=\textwidth]{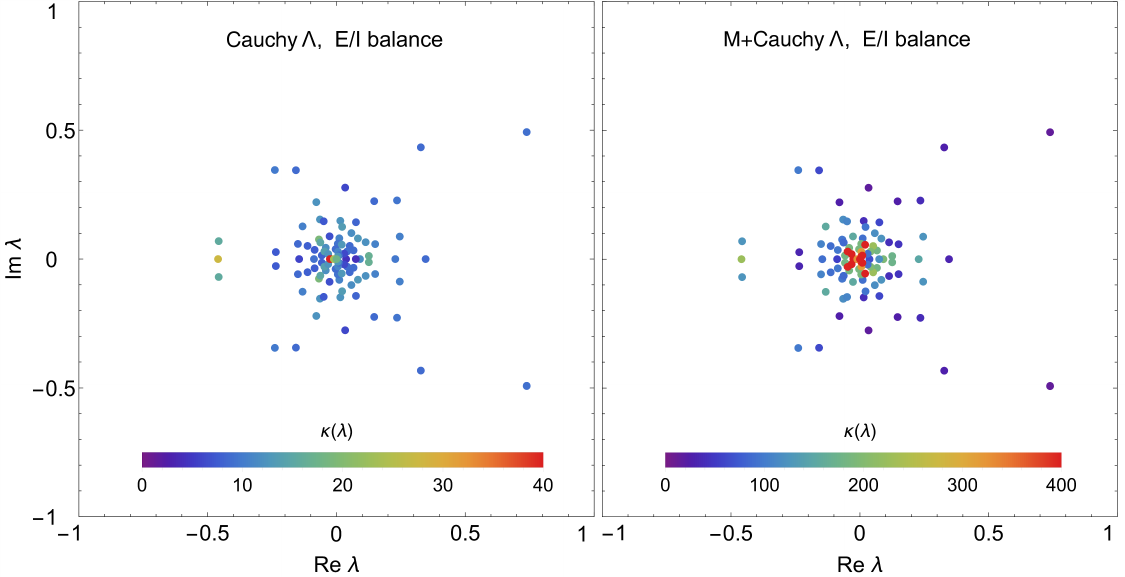} 
\caption{Eigenvalues and their condition numbers for the synaptic matrix, the random part of which is generated from the matrix Cauchy distribution without (left) and with (right) the deterministic connection, $M$, reflecting Dale's principle. Note the increase of condition numbers caused by addition of $M$ (the scale is resized by an order of magnitude). Matrices $M$ and $\Lambda$ are the same as in Fig~\ref{Fig:SpectraComparison}.
\label{Fig:SpectraComparisonCauchy}
}
\end{figure}

\begin{figure}[h]
\includegraphics[width=\textwidth]{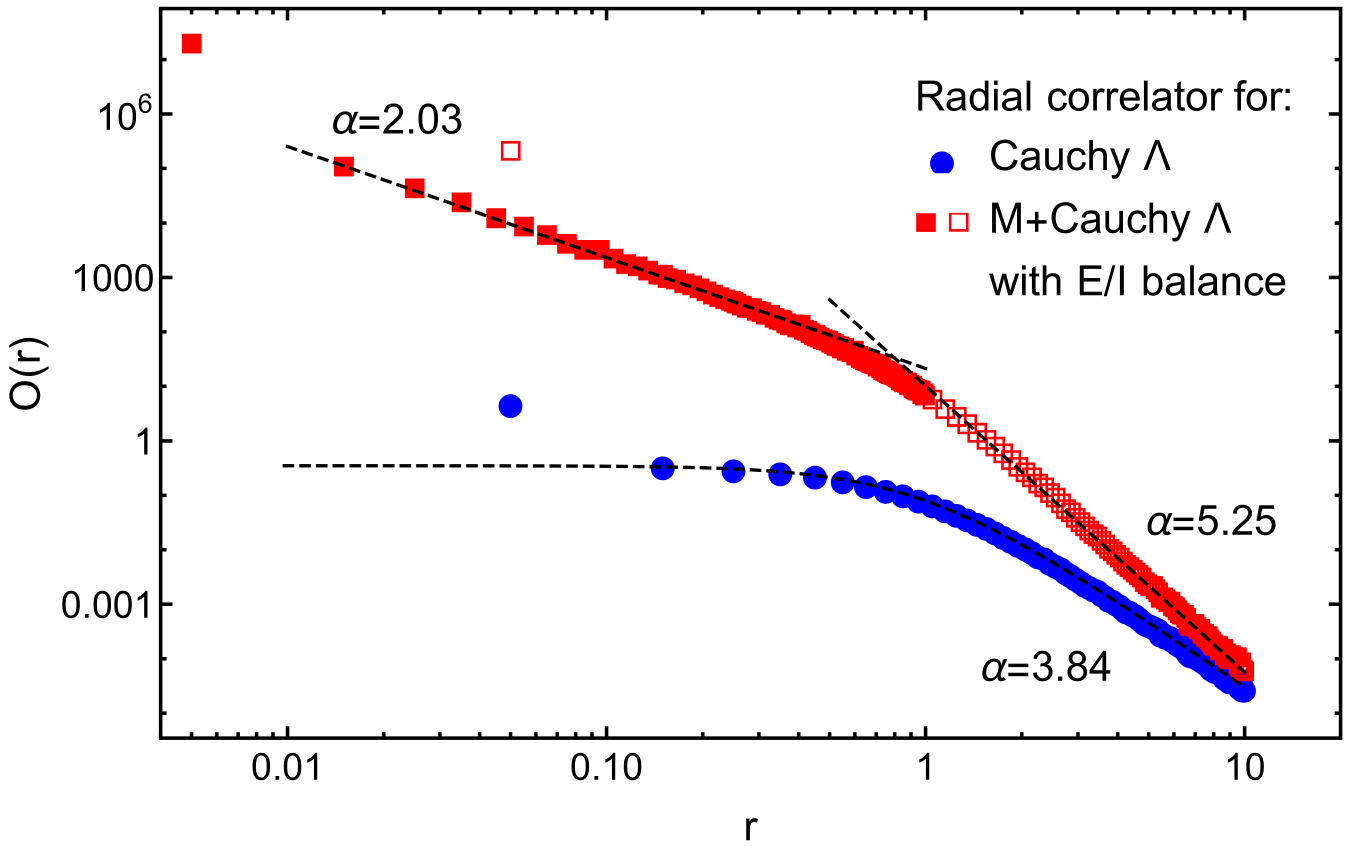} 
\caption{The radial eigenvector correlator for Cauchy synaptic matrices on a log-log scale.
Blue circles approximate the analytic prediction ($\alpha=4$, dashed curve)  for the unperturbed model.
The red slope is the perturbed model, with full squares reflecting the universal inverse squared behavior for small $r$ (exponent $\alpha=2$). Dashed straight lines are numerical fits for small and large $r$.
Matrix size used: $N=500$.\label{Fig:CauchyCorr}
}
\end{figure}

Above conclusions are strengthened  by the similar study based on Cauchy synaptic matrices. Fig~\ref{Fig:FRVCauchy} shows perfect agreement of our predictions with the numerics. Due to the local E/I balance the spectra are unchanged. This does not hold, however, for the squared eigenvalue condition numbers --  they dramatically increase (several orders of magnitude, note the scales in Figs \ref{Fig:SpectraComparisonCauchy} and \ref{Fig:CauchyCorr}). Finally, the unperturbed  eigenvector correlator approaches the predicted slope (compare the predicted slope 4 to the measured 3.84). The perturbed correlator reproduces  small $r$ behavior (compare the predicted exponent 2 to the measured 2.03), whereas large $r$ numerical simulations provide asymptotic  slope 5.25, as compared to the  predicted slope equal to 4. 

\begin{figure}[h]
\includegraphics[width=\textwidth]{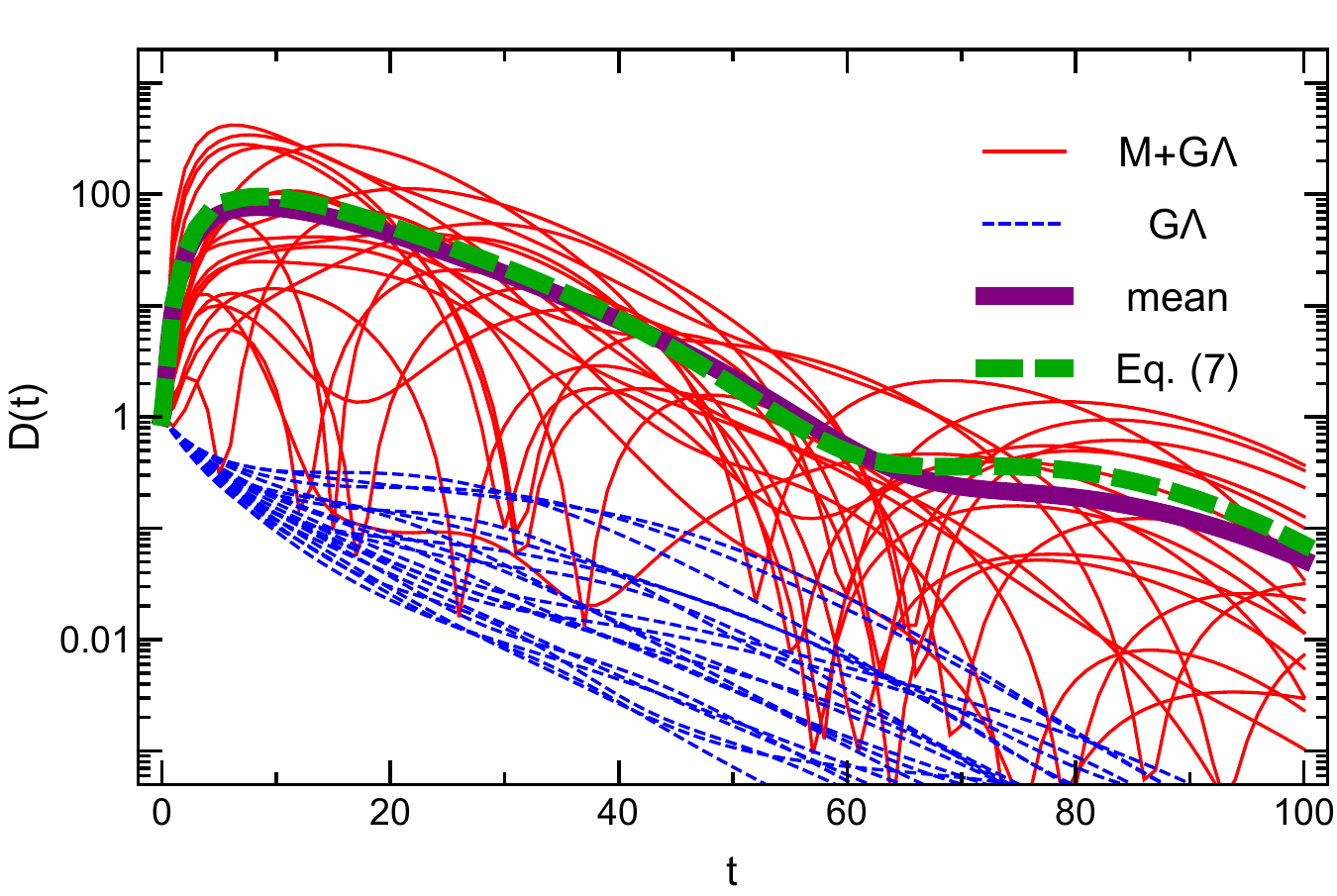} 
\caption{
The squared Euclidean distance from the fixed point in the linearized dynamics of~\citet{WAINRIB}. The presence of $M$ induces transient behavior and strong oscillations. These effects are caused by  strong non-normality. Numerical results were obtained for the minimal Rajan-Abbott model. The matrix is of size $N=100$ with the same parameters as in Fig~\ref{Fig:SpectraComparison}. We chose $\mu=r_{out}+0.02$ to ensure stability. Each blue and red curve corresponds to a single initial condition generated randomly from the set of vectors of unit norm. Solid purple line represents an average of $D(t)$ taken over the presented realizations, while the green dashed line is the theoretical average over all initial conditions, Eq. \eqref{unitary}.
\label{Fig:SquaredNorm}
}
\end{figure}

Although the presence of the matrix $M$ breaks the unitary invariance of the synaptic strength matrix, it is still worth considering the squared norm averaged over initial conditions as a quantity measuring the transient response.
The deterministic connections and the E/I balance cause an increase of all elements of the overlap matrix $O_{ij}$, as Eq~(\ref{eq:OffDiag}) predicts. To elucidate the importance of this fact, we studied the squared norm of the solution to the linearized dynamics (\ref{eq:SquaredNorm}) with $X=W$ and $X=M+W$, where initial conditions were generated randomly from the uniform distribution on the unit sphere. This dynamics is obtained by the linearization of the model considered in~\citep{WAINRIB}. Results presented in Fig~\ref{Fig:SquaredNorm} show that the deterministic connections in the network followed by the E/I balance significantly enhance the norm of the solution and all presented trajectories are transient. This would not be the case if the connections were fully random. Moreover, the strong oscillations of the squared norm indicate  interference between the eigenmodes. It is worth stressing the accuracy of qualitative predictions based on Eq \eqref{unitary} despite the fact that the presence of the matrix $M$ and the E/I balance break the rotational symmetry of the ensemble.

\begin{figure}
\includegraphics[width=1.0\textwidth]{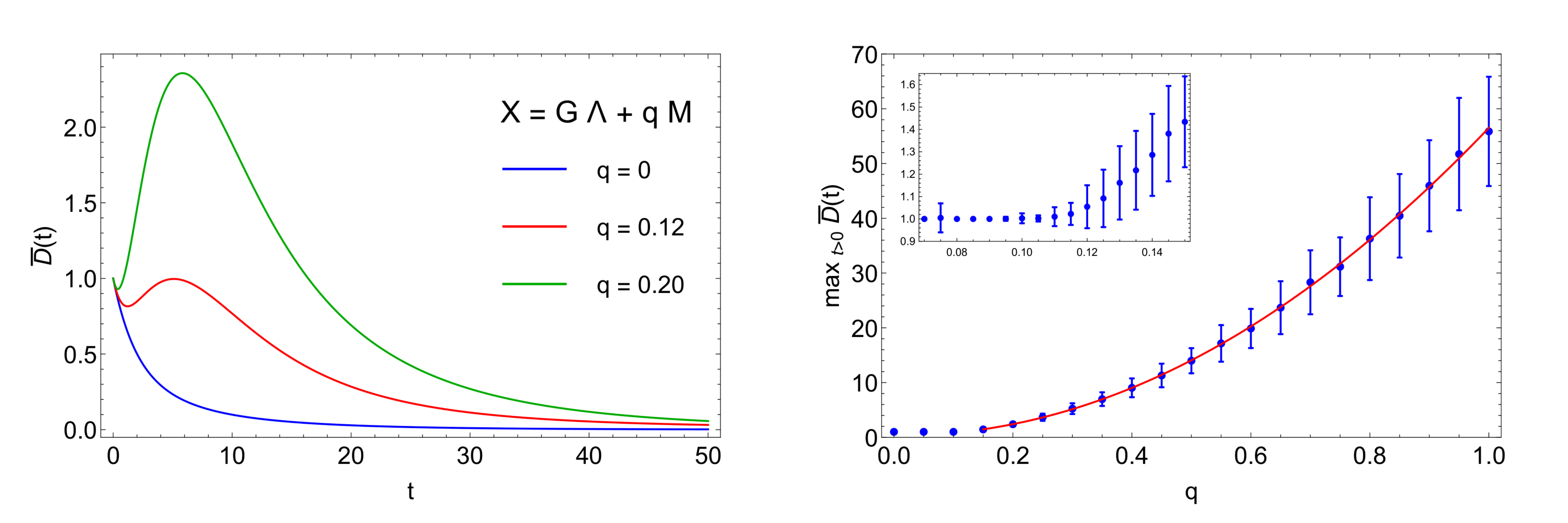}
\caption{(left) The average squared norm, \eqref{unitary}, in the system \eqref{eq:LinearDynamics} with the connectivity matrix $X=G\Lambda+qM$. The parameter $0\leq q\leq 1$ tunes the strength of deterministic connections. We chose $\mu=r_{out}+0.05$, $N=100$ and averaged over 200 realizations of the matrix $X$.
(right) Maximum of the squared norm averaged over initial conditions, Eq \eqref{unitary}, further averaged over 200 realizations of the matrix $X$. Error bars denote standard deviation. The red line depicts the quadratic fit $0.38 - 1.35 q + 57.48 q^2$ for data with $q\geq 0.15$, confirming predictions based on Eq \eqref{eq:OffDiag}. In the inset we show a close-up of the region around $q^*$ where the transition between the constant and quadratic behavior takes place.
\label{Fig:Transients}
}
\end{figure}

To further explore the effect of the matrix $M$ we study the linear dynamics governed by the synaptic connectivity matrix $X=W+qM$ and imposed E/I balance. The parameter $0\leq q \leq 1$ allows one to tune the strength of the deterministic weights and the level of non-normality. For $q=1$ it coincides with the Rajan-Abbott model. Numerical simulations show that for small values of $q$ the squared norm decays monotonically. As $q$ increases, $\bar{D}(t)$ becomes non-monotonic with a local maximum (see Fig~\ref{Fig:Transients}). For a quantitative study, as a measure of the transient amplification we consider the maximum of the squared norm over the entire time span, $\max_{t>0} D(t)$, as a function of $q$. Eq~\eqref{unitary} shows that the transient dynamics is governed by the full overlap matrix and--according to Eq~\eqref{eq:OffDiag} where the matrix $M$ enters twice--we expect that the maximal amplification grows quadratically with $q$. This behavior is verified in Fig~\ref{Fig:Transients} and is true only for $q$ exceeding a certain threshold $q^*$. For $q<q^*$ the transient effects are small and the maximal value is the initial value, 1. 


\begin{figure}[h]
\includegraphics[width=0.99\textwidth]{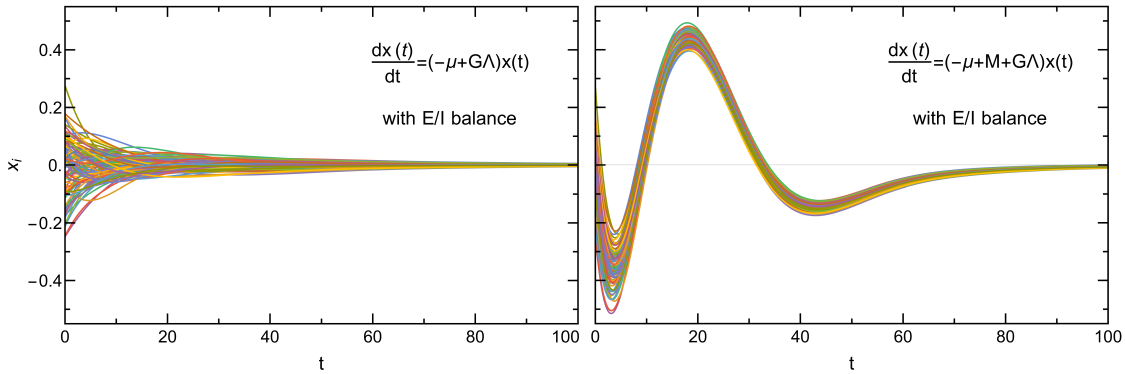}  
\caption{The activity of each neuron in the linearized dynamics. In the right panel we can see the onset of collective dynamics driven by the matrix $M$ and the balance condition. Both simulations started from exactly the same initial condition randomly chosen from the $N$-dimensional hypersphere. Parameters are the same as in Fig~\ref{Fig:SquaredNorm}.
\label{Fig:Signals}
}
\end{figure}

One expects these dramatic effects to be visible in the activity of individual neurons. We therefore studied the temporal dynamics of the components of the vector of neural activities (\ref{eq:FormalSol}) for randomly chosen initial conditions. The results, presented in Fig~\ref{Fig:Signals}, show that in the presence of $M$, the neuronal activity is not only transiently enhanced, but also more synchronized, as observed numerically in the full dynamics by~\citet{WAINRIB}. This effect, which is persistent in the non-linear model, is observed as transient in the linearized dynamics.

Spectra of heavy-tailed random matrices are unbounded and there is a nonzero probability for arbitrarily large eigenvalues. This challenges the model \eqref{eq:LinearDynamics} with fixed $\mu$. We adapt it to the heavy-tailed spectra, noticing that for each realization of the random matrix the corresponding eigenvalues are finite. To ensure the stability of the linear system, we choose $\mu=0.02+\max \re \lambda$ for each realization of randomness. Similarly to the case of the Gaussian disorder we observe an initial growth of $D(t)$, which is two orders of magnitude stronger in the presence of the matrix $M$, followed by relaxation towards the fixed point and oscillations resulting from the interference between eigenmodes. The spectral radius of the Cauchy matrix grows with its size like $\sqrt{N}$~\citep{JQ}, therefore to ensure stability, $\mu$ needs to be of the same order, while for the Gaussian noise $\mu$ is of order one. This difference of scales is equivalent to different time scales of the dynamics, therefore the transient effects last much shorter with Cauchy noise. This effect is further magnified because of the low value of the spectral density for large $r$. Eigenvalues with large real parts are more separated from each other and a single mode quickly dominates Eq~\eqref{unitary}, ending the transient phase.


\section{Discussion and conclusions}
\label{sec:summup}
In this letter we explored the use of FRV in the study of large synaptic interaction  matrices. Besides straightforwardly recovering  known results on the application of random matrices to neural networks, we have addressed the issue of large fluctuations, most probably  very relevant to the dynamics of  learning and memory in biological neural networks \citep{LITERATURE,LITERATURE2}. 
Using  recent results on the properties of eigenvectors in non-normal matrices, we have quantitatively linked  the strength of the fluctuation of the outliers to a certain eigenvector correlator. 
We presented our analysis for the simplest Gaussian case, nevertheless, we also pointed out how one can consider other distributions, e.g. heavy-tailed. The formalism stays the same, but    in the case of more general pdf's (apart form the Cauchy disorder which we solved analytically) one may need to rely on numerical solutions. In the case of heavy tails, one needs to redefine the dynamical model~\eqref{eq:LinearDynamics} in such a way that the fixed point is stable. For finite size of matrices it is possible by adjusting the parameter $\mu$.

Previous works on dynamical random matrices~\citep{OURPRL,OURNPB,GRWAR} 
show that the understanding of the temporal evolution of non-normal matrix models requires considering the entangled dynamics of both eigenvectors and eigenvalues, contrary to the  simple evolution of the spectra of normal matrices, for which the eigenvectors decouple in the presence of the spectral evolution.

The synaptic strengths of real neuronal networks are not static~\citep{KANDEL}. Neural activity itself, in the course of time, allows neurons to form new connections, strengthening or weakening the existing synapses.
This synaptic plasticity, on which biological learning is based, is not captured in many models. Nonetheless, the change of the synaptic strengths in a short time interval can be treated as a small additive perturbation of the initial matrix. This results in reorganization of the spectrum on a complex plane. 

Our results indicate that for balanced networks the sensitivity of eigenvalues to additive perturbation is dramatic and increases several orders of magnitude in the networks with a heavy-tailed spectrum of adjacency matrices (small worlds).
Since it is  commonly accepted that spike-timing-dependent plasticity in small-world networks is a hypothetical learning mechanism~(for a recent experimental study see~\citep{KIMLIN}), one may worry how synchronization of the network is possible at all. 
We emphasize here that the E/I balance is put into this model by hand. In the real brain the E/I balance is maintained on the scale of hundreds of milliseconds~\citep{EXPBALANCED3}, and periods during which the balance is violated are not longer than few milliseconds~\citep{EXPBALANCED,EXPBALANCED2}. More complete models of neural networks must incorporate the E/I balance as a dynamical process. 

Networks adapting to the changing external conditions may change their structure in a controlled way. The high sensitivity of eigenvalues to these changes in this case might be desired, because it can  facilitate the adaptation. We hypothesize that such high sensitivity in the models with dynamical E/I balance can emerge through a process a kind of self-regulated criticality~\citep{DANTE}. Although the specifics of such process are not certain yet, there is evidence both empirical~\citep{TURRIGIANO,LIU,PU,SHEW}  as well as theoretical~\citep{LEVINA,MAGNASCO,SCHNEIDMAN} of its plausibility. In addition, the connection of the E/I balance with criticality was already observed at the level of neuronal avalanche analysis in EEG or MEG data~\citep{CRITICALEEG}.

Since the balance condition leads to a dramatic increase of eigenvector overlaps -- conditioning the spectra -- which further take crucial part in driving mechanisms of temporal evolution of the networks,  one needs a powerful, stabilizing mechanism preventing the transition to the chaotic behavior in the full nonlinear dynamics. 

We envision  one a priori mechanism,  which can tame such a behavior -- it is the transient behavior. This conclusion is consistent with the model of~\citet{WAINRIB} for non-normal  balanced networks,  who have observed synchronization inexplicable  by solely spectral properties of the networks. Transient behavior means that even stable trajectories may initially diverge before  reaching the fixed point for long times. This implies that transient behavior is complementary to the stability analysis and may signal non-linear features already on the linear level~\citep{GRELA}. Since analytic tools allowing the study of transient behavior for balanced networks are still missing,  we have performed  sample simulations for Gaussian networks. Results are shown in Fig~\ref{Fig:SquaredNorm}-\ref{Fig:Signals}. These simulation confirm  qualitatively the presence of transient behavior. 

 They raise, however, more quantitative questions: what are the statistical  features of transient behavior in balanced neuronal networks? How do the effects of transient behavior scale  with the size of the network? What are the time-scales in the transient behavior? How does the transient behavior depend on the type of  adjacency matrix? 
 We hope to provide some analytic answers to  these questions in the sequel to this work. 
Last but not least,  considering the success of FRV analysis in a variety of disciplines, we hope that the ideas presented in this  paper may trigger more interdisciplinary interactions in the area of brain studies.

\subsection*{Acknowledgments}
The research was supported by the MAESTRO DEC-2011/02/A/ST1/00119 grant of the National Center of Science. WT also appreciates the financial support from the Polish Ministry of Science and Higher Education through ``Diamond Grant'' 0225/DIA/2015/44 and the scholarship of Marian Smoluchowski Research Consortium Matter Energy Future from KNOW funding. The authors  thank Jacek Grela and Piotr Warcho\l{} for discussions  and critical remarks, and Janina Krzysiak for carefully reading the manuscript. 

\section*{Appendix A: A guide through free random variables}
\label{Sec:app1}
  Free random variables can be viewed as a probability theory, where the basic random variable is represented by an infinite matrix. 
  It is therefore most convenient to explain the cornerstones of the theory of free probability  using  concepts from the classical theory of probability (CTP). For the more detailed treatment of the problem, we refer to~\citep{MINGOSPEICHER}.
 
  Let us consider the following problem. We have two random variables $x_1$ and $x_2$ drawn from independent probability distributions $p_1(x_1)$ and $p_2(x_2)$. The distribution of the random variable $s$ being the sum  of $x_1$ and $x_2$ reads therefore 
  \be
  p(s)=\int dx_1 dx_2 p_1(x_1)p_2(x_2) \delta(s-(x_1+x_2))=\int dx p_1(x)p_2(s-x).
  \ee
  One can easily unravel the convolution using the Fourier transform (characteristic function). 
  Then $\hat{p}(k) \equiv \int p(s) e^{iks}ds =\hat{p}_1(k)\hat{p}_2(k)$, where $\hat{p}_i(k)$ are Fourier transforms corresponding to the original densities $p_i(x)$. 
Note that a characteristic function generates moments of the respective distribution. We can further simplify the problem if instead of characteristic functions we consider their natural logarithms $\phi(k) \equiv \ln \hat{p}(k)$. Then we get the addition law, which linearizes the convolution
  \be
  \phi_{1+2}(k)=\phi_1(k) +\phi_2(k).
  \label{classadd}
  \ee
  Since $\phi$ is another generating function -- this time for cumulants of the distribution -- the above relation means additivity of the corresponding cumulants. 
  The algorithm of convolution is therefore straightforward.  First, knowing $p_i(x)$, we construct $\phi_i(k)$. Then we perform the addition 
  law ~(\ref{classadd}). Finally, we reconstruct 
$p_{1+2}(s)$  from  $\phi_{1+2}(k)$, performing the first step in reversed order. 
A pedagogical and simple example is represented by the convolution of two independent Gaussian distributions $N_1(0, \sigma_1^2)$ and $N_2(0, \sigma_2^2)$ .  The first step shows that in both cases only one cumulant, i.e. the second one, $\kappa_2=\sigma^2$,  is non-vanishing. The addition law and the last step of the logarithm immediately lead to the the resulting distribution, which is also Gaussian, 
$N_{1+2} (0, \sigma^2=\sigma_1^2+\sigma_2^2)$. 

 In free probability, the  notion of independence  is replaced  by the notion of freeness. Two large (infinite) matrices are mutually free if  their eigenvectors are maximally  decorrelated, e.g.,  matrices $X$ and $UYU^{\dagger}$, where $U$ is the Haar measure, are free. 
 
 The role of the characteristic function is played by the complex valued Green's function
 \be
 G_X(z)=\int \frac{\rho_X(\lambda)}{z-\lambda}d\lambda,
 \ee
where $\rho_ X(\lambda)$ is the average spectral density of the matrix $X$, playing here the role of the probability density function in CTP. Indeed, expanding $G_X(z)$ around $z=\infty$ we get spectral moments $m_k^{(X)}=\int \lambda^k\rho_X(\lambda)d\lambda$.  Note, that knowing $G_X(z)$ we  can easily reconstruct $\rho_X(\lambda)$. Indeed, 
\be
\begin{aligned}
-\frac{1}{\pi} \lim_{\epsilon \rightarrow 0} \Im G(z)|_{z=\lambda^{'} +i\epsilon} =&
\lim_{\epsilon \rightarrow 0}\int \rho(\lambda) \frac{1}{\pi} \frac{\epsilon}{({\lambda^{'}-\lambda)}^2 +\epsilon^2} d\lambda =\\  
=&
\int \rho(\lambda)\delta(\lambda-\lambda^{'})d\lambda= \rho(\lambda^{'}).
\end{aligned}
\ee
The role of the generating function for free cumulants is played by the so-called R-transform, $R(z)=\sum_{k=1}^{\infty}\kappa_kz^{k-1}$. 
The crucial relation between $R(z)$  and $G(z)$ reads $ R(G(z))+\frac{1}{G(z)}=z$ or $G(R(z)+1/z))=z$, i.e., the function $R(z)+\frac{1}{z}$ is the functional inverse of the Green's function. 
Let us come back to the problem of addition.  Imagine we have now the spectral measures $\rho_{X_i}(\lambda)$, corresponding to two matricial ensembles  with the measures $P(X_i)dX_i$, where $i=1,2$.  We are now asking what the spectral density of the ensemble $X_{1+2}=X_1+X_2$ is. This is a highly non-trivial and non-linear problem, since $X_1$ and $X_2$ do not commute, but free calculus allows to solve this case in full analogy to CTP.  The algorithm is as follows: first,  from $\rho_i$ corresponding  to $X_i$ we construct matching $G_i(z)$ and $R_i(z)$. Then 
\be 
R_{X_1+X_2}=R_1(z)+R_2(z),
\ee
which supersedes Eq~(\ref{classadd}). 
Finally, we proceed in reverse order, reconstructing from $R_{X_1+X_2}(z)$ the Green's function $G_{X_1+X_2}(z)$, and finally the spectral density
$\rho_{X_1+X_2}(\lambda)$. 
As an example, we consider the ``Gaussian" distribution in free theory, i.e., the spectral distribution where the only non-vanishing cumulant is the variance $\sigma^2$. Thus $R(z)=\sigma^2 z$.  Reconstructing Green's function gives $\sigma^2 G+1/G=z$, with the solution $G(z)=\frac{1}{2\sigma^2} (z-\sqrt{z^2-4\sigma^2})$.  Taking the imaginary part, we reconstruct the celebrated Wigner semicircle
$\rho(\lambda)=\frac{1}{2\pi \sigma^2}\sqrt{4\sigma^2-\lambda^2}$. We see that the addition algorithm for two free Wigner semicircles mimics precisely the addition algorithm of two Gaussians. 

Similarly to addition, one can consider multiplication laws for random variables $x_1 \cdot x_2$. In CTP, such a problem is unravelled 
with the help of the Mellin transform~\citep[see e.g.,][]{EPSTEIN}. In  free calculus,  the role of the Mellin transform is played by the S-transform, related to the R-transform by $S_X(z)R_X(zS_X(z))=1$. The multiplication law reads 
\be 
S_{X_1X_2}(z)=S_{X_1}(z) \cdot S_{X_2}(z)
\ee
and the algorithm for multiplication follows the algorithm for addition. 
However, one should be aware that the product of two symmetric (Hermitian) matrices may be non-symmetric (non-Hermitian). In such a case, the eigenvalues can appear  on the whole  complex plane, and the methods of $R(z)$ and $S(z)$ transforms, based on analyticity,  require substantial modifications. 
Luckily, there exists one powerful case, governed by the Haagerup-Larsen theorem (known also as the ``single ring" theorem), when analytic methods  hold for complex spectra. If the complex matrix $X$ can be decomposed as $X=PU$, where $P$ is positive, $U$ is Haar-measured and $P$ and $U$ are mutually free, the spectrum on the complex plane has a polar symmetry and the radial distribution  can be easily read out  from the singular values of $X$, i.e., the real eigenvalues of $X^{\dagger}X$.  In free probability theory, such ensembles are known as R-diagonal.  To infer the information about the spectra and some correlations between left and right eigenvectors one needs only the explicit form of $S_{X^{\dagger}X}(z)$. 
In the case of the Ginibre ensemble $G$ (i.e., where $G_{ij}$ are drawn either from real or complex Gaussian distributions),  this is particularly easy, since matrix $G^{\dagger}G$ is known as a Wishart ensemble.  To avoid obscure mathematics,  let us recall that the  Wishart ensemble is a free analogue  of the Poisson distribution from classical probability~\citep{Voiculescu}. This implies that all cumulants are the same, and if  normalized to  1 for convenience, its R transform is  by definition  $R_{G^{\dagger}G}(z)=\sum_{i=1}^{\infty}  z^{i-1}=\frac{1}{1-z}$.  Using the above-mentioned functional relation between R and S transforms we arrive at $S_{G^{\dagger} G}=\frac{1}{1+z}$.  Similar techniques can be applied for generic randomness in Rajan-Abbott type models, as we show below. 

\section*{Appendix B: The Rajan-Abbott  model with Gaussian noise}
\label{Sec:RAGauss}

We use the theorem from free probability, which states that the product of an R-diagonal operator with any operator is R-diagonal~\citep{RDIAGPRODUCT}, therefore $W$ is subject to the Haagerup-Larsen theorem.  Then $W^{\dagger}W=\Lambda G^{\dagger}G \Lambda \stackrel{\mathrm{Tr}}{=} G^{\dagger}G  \Lambda^2$, where the last equation expresses the fact that the spectral properties are invariant under the cyclic permutations of matrices under the trace. The Green's function (resolvent) for $\Lambda^2$ reads therefore
\be
G_{\Lambda^2}(z)=\sum_{i=1}^{m}\frac{f_i}{z-\sigma^2_i}.
\label{Green}
\ee
Substituting $z \rightarrow R_{\Lambda^2}(z) +\frac{1}{z}$ in Eq~(\ref{Green}) and using the fundamental FRV relation $G(R(z) +\frac{1}{z})=z$  we arrive at 
\be
1=\sum_{i=1}^{m} \frac{f_i}{zR_{\Lambda^2}(z) -z\sigma^2_i+1}.
\label{aux1}
\ee
Now we replace in Eq~(\ref{aux1}) $z \rightarrow tS_{\Lambda^2}(t)$ and using the relation between $S$ and $R$ transforms we arrive at 
\be 
1=\sum_{i=1}^{m} \frac{f_i}{1+t-\sigma_i^2 tS_{\Lambda^2}(t)}.
\label{main1}
\ee
We note that $\frac{1}{1+t}$ is the S-transform for the Wishart ensemble (calculated above), and the multiplication law gives us the final S-transform for $W^{\dagger}W$, 
i.e., $\frac{1}{1+t} S_{\Lambda^2}(t)= S_{G^{\dagger}G}(t) S_{\Lambda^2}(t)=S_{W^{\dagger}W}(t)$, so we arrive at
\be
 1+t= \sum_{i=1}^{m} \frac{f_i}{1-\sigma_i^2 tS_{W^{\dagger}W}(t)}.
 \ee
 In the last step  we substitute $t \rightarrow F(r)-1$ and use the Haagerup-Larsen theorem, arriving at 
 \be
 F(r)=\sum_{i=1}^{m} \frac{f_i}{1-\sigma_i^2(F(r)-1)/r^2}.
 \ee
 Subtracting $1=\sum_i f_i$ from both sides, we simplify it to
 \be
 1=\sum_{i=1}^{m} \frac{f_i\sigma_i^2}{r^2-\sigma_i^2(F(r)-1)}.
 \label{final}
 \ee

\section*{Appendix C: The Rajan-Abbott model with Cauchy noise}
\label{Sec:RACauchy}

FRV calculus is a powerful technique and the range of its applications is not confined to the basin of attraction of the Gaussian type. In particular, for random matrices $X$ belonging to the free   Lévy class (spectral density decays like $1/\lambda^{\alpha-1}$), the S-transform  for the Wishart-Lévy matrix 
$X^{\dagger}X$ reads $S_{X^{\dagger}X}(t)=\frac{1}{t(1+t)}{\left( \frac{t}{b}\right)}^{t/\alpha}$, with $b= \exp [i \pi (\alpha/2-1)]$ ~\citep{NOWAKFIN}. 
The stability index $\alpha=2$ reproduces the Gaussian case,  but a simple form can be obtained also for the Cauchy disorder $\alpha=1$. In this case 
$S_{X^{\dagger}X}(t)=-\frac{t}{1+t}$, and when applied to Eq~(\ref{main1}), yields
\be
1+t=\sum_{i=1}^{m} \frac{f_i}{1+S_{W^{\dagger}W}(t)\sigma_i^2}.
\ee
The final substitution $t \rightarrow F(r)-1$  and the use of the Haagerup-Larsen theorem gives an explicit, linear equation for arbitrary number of types of neurons
\be
 F(r)=  \sum_{i=1}^{m} \frac{f_i}{1+\sigma_i^2/r^2}.
 \ee
 Contrary to the previous case, the spectrum is unbounded and stretches up to infinity. Explicitly, the spectral density and the eigenvector correlator read
 \be
 \rho(r)&=&\frac{1}{2\pi r}\frac{dF(r)}{dr}=\frac{1}{\pi} \sum_{i=1}^{m} \frac{f_i \sigma_i^2}{{(r^2+\sigma_i^2)}^2}, \\
 O(r) &=& \frac{1}{\pi r^2} F(r)(1-F(r))= \frac{1}{\pi} \sum_{i=1}^{m}\frac{f_i}{r^2+\sigma_i^2} \sum_{j=1}^{m} \frac{f_j\sigma_j^2}{r^2+\sigma_j^2}.
 \label{Cauchyresults1}
 \ee
 In the case of arbitrary $\alpha$, resulting transcendental equations can be easily solved numerically.
Other types  of  neural network randomness can also be modeled, e.g. by considering Student-Fisher spectral distributions. 


\end{document}